\documentclass[aps,twocolumn,superscriptaddress,preprintnumbers,floats]{revtex4}
\usepackage[colorlinks,citecolor=blue,anchorcolor=red,menucolor=red,linkcolor=red,filecolor=red,urlcolor=blue,frenchlinks=red]{hyperref}
\usepackage{amsfonts}
\usepackage{amsmath}
\usepackage{amssymb}
\usepackage{CJKutf8}
\usepackage{color}
\usepackage{comment}
\usepackage{epsfig}
\usepackage{epstopdf}
\usepackage{float}
\usepackage{graphicx,booktabs}
\usepackage{indentfirst}
\usepackage{longtable,lscape}
\usepackage{mathrsfs}
\usepackage{mathtools}
\usepackage{morefloats}
\usepackage{pifont}
\usepackage{txfonts}
\begin{document}

\title{Fully-strange tetraquarks: fall-apart decays and experimental candidates}

\author{Feng-Xiao Liu}
\email[E-mail: ]{lfx@usc.edu.cn}
\affiliation{Institute of High Energy Physics, Chinese Academy of Sciences, Beijing 100049, China}
\affiliation{School of Nuclear Science and Technology, University of South China, 421001 Hengyang, China}

\author{Xian-Hui Zhong}
\email[E-mail: ]{zhongxh@hunnu.edu.cn}
\affiliation{Department of Physics, Hunan Normal University, and Key Laboratory of Low-Dimensional Quantum Structures and Quantum Control of Ministry of Education, Changsha 410081, China}
\affiliation{Synergetic Innovation Center for Quantum Effects and Applications (SICQEA), Hunan Normal University, Changsha 410081, China}

\author{Qiang Zhao}
\email[E-mail: ]{zhaoq@ihep.ac.cn}
\affiliation{Institute of High Energy Physics, Chinese Academy of Sciences, Beijing 100049, China}
\affiliation{University of Chinese Academy of Sciences, Beijing 100049, China}
\affiliation{Center for High Energy Physics, Henan Academy of Sciences, Zhengzhou 450046, China}

\begin{abstract}

We present a systematic analysis of the fall-apart decays for the $1S$, $1P$, and $2S$-wave fully-strange tetraquark states. It shows that most of the fully-strange tetraquark states have a relatively narrow fall-apart decay width of $\mathcal{O}(10)$ MeV. The newly observed axial-vector state $X(2300)$ at BESIII may favor the low-lying $1S$-wave $1^{+-}$ state $T_{(4s)1^{+-}}(2323)$, while the $X(2500)$ resonance observed in the earlier BESIII experiment may favor the low-lying $1P$-wave $0^{-+}$ state $T_{(4s)0^{-+}}(2481)$. Some fully-strange tetraquark states predicted in theory can be searched for in their dominant fall-apart decay channels in experiment, such as $\phi\phi$, $\phi\phi(1680)$, $\eta^{(\prime)}\phi$, $\eta^{(\prime)}h_1(1415)$, and $\phi f_2^{\prime}(1525)$, to which they have relatively large couplings. 

\end{abstract}


\maketitle

\section{Introduction}\label{sec:intro}

Searching for genuine exotic hadrons beyond the conventional quark model has been one of the most important initiatives since the establishment of the quark model in 1964~\cite{Gell-Mann:1964ewy,Zweig:1964ruk}. Recently, an exotic resonance $X(6900)$ was observed in the di-$J/\psi$ invariant mass spectrum by the LHCb collaboration~\cite{LHCb:2020bwg}, which was also confirmed by the CMS and ATLAS collaboration three years later~\cite{CMS:2023owd,ATLAS:2023bft}. In addition, in the lower-mass region the CMS measurements show that exotic resonances $X(6600)$ and $X(7100)$ can also be established in the di-$J/\psi$ spectrum. These exotic resonances may be evidence for genuine fully-charmed tetraquark states. As an analogy, one may expect to observe stable fully-strange tetraquark $T_{ss\bar{s}\bar{s}}$ states as well in experiments though some ambiguities may be inevitable due to the light mass of the strange quark.

Over the past twenty years, several $T_{ss\bar{s}\bar{s}}$ candidates have been observed in experiments. For example, the $f_{0}(2200)$ and $f_{2}(2340)$ resonances listed in the Review of Particle Physics (RPP)~\cite{ParticleDataGroup:2024cfk} may favor the $1S$-wave $T_{ss\bar{s}\bar{s}}$ states according to the mass analysis in the literature~\cite{Liu:2020lpw,Ebert:2008id}. In 2016, the BESIII collaboration observed a new resonance $X(2500)$ in the di-$\phi$ invariant mass spectrum of the $J/\psi\to\gamma\phi\phi$ process~\cite{BESIII:2016qzq}. The $X(2500)$ may be explained as the $1P$-wave $T_{ss\bar{s}\bar{s}}$ state with $J^{PC}=0^{-+}$ from the point view of the mass~\cite{Liu:2020lpw,Dong:2020okt}. Very recently, a new axial-vector state with $J^{PC}=1^{+-}$, denoted as $X(2300)$, was observed in the $\phi\eta$ and $\phi\eta^{\prime}$ invariant mass spectra by the BESIII collaboration~\cite{BESIII:2024nhv}. With the interpretation of conventional $s\bar{s}$ state $h_1(3P)$, the observed mass is about 100 MeV lower than the theoretical predictions~\cite{Li:2020xzs}. However, it is interesting to find that the observed mass of $X(2300)$ is consistent with our prediction for the $T_{(ss\bar{s}\bar{s})1^{+-}}(2323)$~\cite{Liu:2020lpw}. To further confirm these candidates as $T_{ss\bar{s}\bar{s}}$ states, in addition to the mass spectrum, a systematic study of the fall-apart decay properties is also crucial.

In the present work, we will further systematically explore the fall-apart decay mechanism of the $T_{ss\bar{s}\bar{s}}$ states by combining our previous predictions of the mass spectrum~\cite{Liu:2020lpw}. Although there are many studies on the $T_{ss\bar{s}\bar{s}}$ mass spectrum from various models and approaches, such as QCD sum rules~\cite{Wang:2006ri,Chen:2008ej,Chen:2018kuu,Wang:2019nln,Cui:2019roq,Azizi:2019ecm,Dong:2020okt,Su:2022eun,Jiang:2023atq}, quark models~\cite{Ebert:2008id,Drenska:2008gr,Lu:2019ira,Deng:2010zzd}, detailed calculations of the decay properties of $T_{ss\bar{s}\bar{s}}$ are still limited. So far, the studies on the decay mechanisms are still limited in the literature. Some efforts are made on the fall-apart decays~\cite{Jiang:2023atq,Drenska:2008gr} and radiative transitions~\cite{Lu:2019ira} based on different solutions for the wavefunctions. In this work we will focus on the $1S$-, $1P$-, and $2S$-wave $T_{ss\bar{s}\bar{s}}$ states, and look for candidates for the $T_{ss\bar{s}\bar{s}}$ states from the decay properties. In particular, we will show that some decay channels may be useful for further identifying the $T_{ss\bar{s}\bar{s}}$ states in experiments.

The paper is organized as follows. In Sec.~\ref{sec:framework}, the theoretical framework is introduced. In Sec.~\ref{sec:results}, numerical results concerning the properties of the fully strange states are presented and discussed. Finally, a brief summary is provided in Sec.~\ref{sec:sum}.

\section{Theoretical framework}\label{sec:framework}

\begin{figure}[htbp]
\centering
\includegraphics[width=0.90\columnwidth]{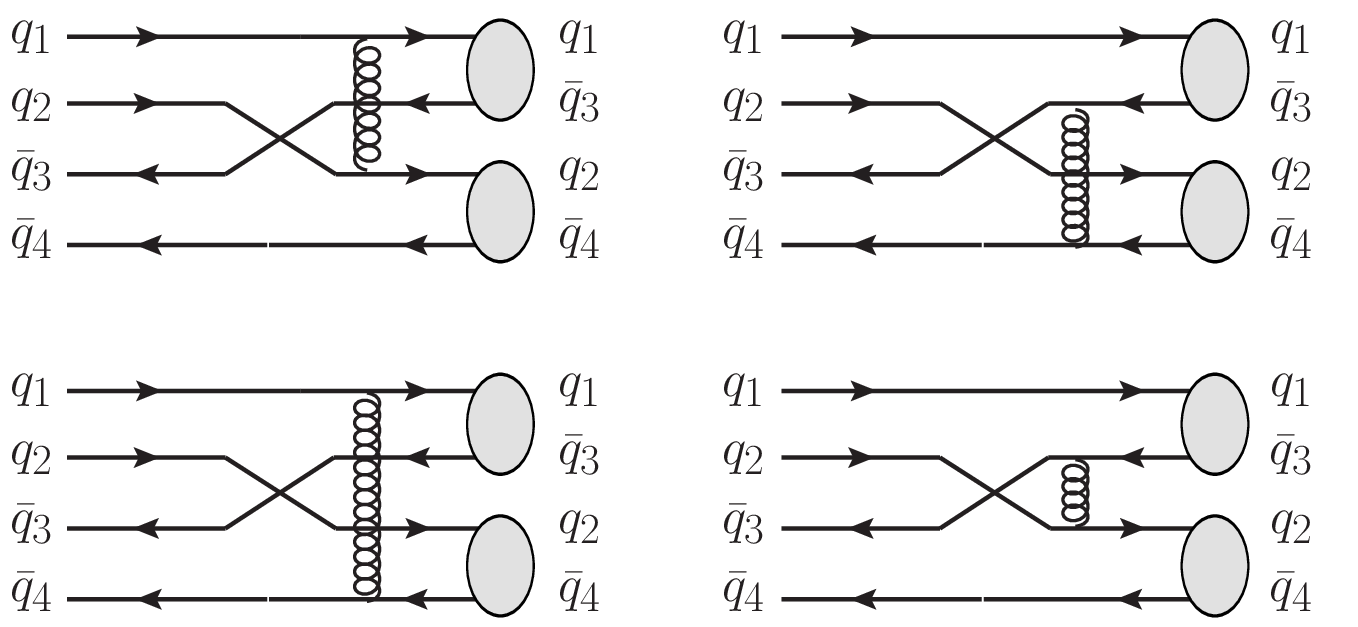}
\caption{The fall-apart decays of
a $T_{ss\bar{s}\bar{s}}$ state induced by the interactions
$V_{ij}$ ($ij\neq 13,24$ or $ij\neq 14,23$) between inner quarks of final hadrons $B$ and $C$.}
\label{fig:decay}
\end{figure}

In our previous work~\cite{Liu:2020lpw}, the mass spectra of the $1S$-, $1P$-, and $2S$-wave $T_{ss\bar{s}\bar{s}}$ states were systematically studied within a nonrelativistic quark model. By using the masses and wave functions obtained from Ref.~\cite{Liu:2020lpw}, in this work we further evaluate the fall-apart decays of the $T_{ss\bar{s}\bar{s}}$ states within a quark-exchange model~\cite{Barnes:2000hu}. In this model, the quark-quark interactions as part of the Hamiltonian are considered to be the sources of the fall-apart decays of multiquark states via the quark rearrangement, which is shown in Fig.~\ref{fig:decay}.

For the decay process $A\to BC$, the decay amplitude $\mathcal{M}(A\to BC)$ is
described by
\begin{eqnarray}
\mathcal{M}(A\to BC)=-\sqrt{(2\pi)^3}\sqrt{8M_AE_BE_C}\left\langle BC |\sum_{i<j} V_{ij}| A \right\rangle,
\end{eqnarray}
where $A$ stands for the initial multiquark state, $BC$ stands for the final hadron pair. $M_A$ is the mass of the initial state, while $E_B$ and $E_C$ are the energies of the final states $B$ and $C$ in the initial-hadron-rest system, respectively. While $V_{ij}$ stands for the interactions between the inner quarks of final hadrons $B$ and $C$ (note that $ij\neq13,24$ or $ij\neq14,23$), they are taken the same form as that of the nonrelativistic potential model adopted in Ref.~\cite{Liu:2020lpw}. The details of the potentials $V_{ij}$ and the parameters were given in Eq.(2) and Table I of Ref.~\cite{Liu:2020lpw}, respectively.

Then, the partial decay width $\Gamma$ for the decay process $A\to BC$ is given by
\begin{eqnarray}
\Gamma & = & \frac{1}{s!}\frac{1}{2J_{A}+1}\frac{|\boldsymbol{p}|}{8\pi M_{A}^{2}}\left|\mathcal{M}(A\to BC)\right|^{2},
\end{eqnarray}
$|\boldsymbol{p}|$ stands for the magnitude of the three momentum for the final states $B$ and $C$ in the rest system of the initial $A$ state. The term $1/s!$ represents a statistical factor that accounts for the indistinguishability of particles. In scenarios where the final state contains two or more identical particles, it is necessary to divide by the number of permutations among these particles to avoid overcounting, as they are indistinguishable from one another. This phenomenological quark-exchange model has achieved a good description of the low-energy $S$-wave phase shift for the $I=2$ $\pi\pi$ scattering at the quark level~\cite{Barnes:2000hu,Barnes:1991em}. Recently, this model has also been extended to study the fall-apart decays of multiquark states in the literature~\cite{Wang:2019spc,Xiao:2019spy,Wang:2020prk,Han:2022fup,Liu:2024fnh,Liu:2022hbk,liu:2020eha}, and a lot of inspiring results are obtained.

In the present work, the masses and wave functions of the initial $T_{ss\bar{s}\bar{s}}$ states are adopted from the numerical results obtained in our previous work~\cite{Liu:2020lpw}. For the final $B$ and $C$ meson states, for simplicity, the wave functions are adopted in a single harmonic oscillator (SHO) form. Their SHO parameters are determined by fitting the root mean square radii of the $s\bar{s}$ states, which are obtained from the potential model calculations using the same Hamiltonian as in Ref.~\cite{Li:2020xzs}. Our determined SHO parameters for the final meson states are collected in Table~\ref{parameters}. For the well established meson states, the masses are adopted from the PDG averaged values~\cite{ParticleDataGroup:2022pth}, otherwise, the meson masses are adopted the potential model predictions. The masses of the final meson states are collected in Table~\ref{parameters} as well. It should be mentioned that in our analysis, the $\eta$ and $\eta^{\prime}$ are treated as mixed states between $s\bar{s}$ and $n\bar{n}$ ($n\bar{n}\equiv (u\bar{u}+d\bar{d})/\sqrt{2}$), i.e.,
\begin{align}
\eta =n\bar{n} \cos \phi_P -s\bar{s}\sin \phi_P ,\ \ \eta^{\prime}= n\bar{n}\sin \phi_P +s\bar{s}\cos\phi_P,
\end{align}
where the mixed angle is taken to be $\phi_P=41.2^\circ$, which is determined by fitting the $\eta^{\prime}$ photo-production data in Ref.~\cite{Zhong:2011ht}.

\begin{table}[htp]
\caption{The masses (MeV) and effective SHO parameters $\beta$ (MeV) of the final meson states.
The missing $2^{1}S_{0}$ $s\bar{s}$ state is denoted by $\eta'(2S)$.
There is still controversy for both the $f_{0}\left(1370\right)$ and $f_{1}\left(1420\right)$
resonances as the assignments of $s\bar{s}$ states, which is marked with a superscript ``?". }\label{parameters}
\begin{tabular}{ccccc}
\hline
\hline
State ~~ & Meson ~~& $M_{exp}$~\cite{ParticleDataGroup:2022pth} &~~$M_{th}$ ~\cite{Li:2020xzs}&~~$\beta$ \tabularnewline
\hline
$1^{1}S_{0}$ & $\eta$ / $\eta^{\prime}$ &~~548 / 958~~ & $\cdot\cdot\cdot$ / 797& 466 \tabularnewline
$1^{3}S_{1}$ & $\phi\left(1020\right)$ & 1019 &1017 & 388 \tabularnewline
$2^{1}S_{0}$ & $\eta'(2S)$ & $\cdot\cdot\cdot$ &1619& 367 \tabularnewline
$2^{3}S_{1}$ & $\phi\left(1680\right)$ &$1680$& 1699 & 330\tabularnewline
$1^{3}P_{0}$ & $f_{0}\left(1370\right)?$& $1400\pm 100$ & 1373 & 365 \tabularnewline
$1^{1}P_{1}$ & $h_{1}\left(1415\right)$&1409 & 1462 & 334 \tabularnewline
$1^{3}P_{1}$ & $f_{1}\left(1420\right)?$&1428 & 1492 & 327 \tabularnewline
$1^{3}P_{2}$ & $f_{2}^{\prime}\left(1525\right)$& 1517 & 1513 & 308 \tabularnewline
\hline \hline
\end{tabular}
\end{table}

\begin{table*}\centering
\caption{Partial fall-apart decay widths (MeV) of the $T_{ss\bar{s}\bar{s}}$ states. In this table, 'o' denotes calculated values less than 0.05~MeV, and '-' denotes a forbidden decay channel. The $2^1S_0$ and $2^3S_1$ $s\bar{s}$ states involving in the fall-apart decay channels are denoted by $\eta^{\prime}(2S)$ and $\phi(2S)$, respectively. While the $1P$ $s\bar{s}$ states with $J^{PC}=0^{++},1^{+-},1^{++},2^{++}$ are denoted by $f_{0}$, $h_{1}$, $f_{1}$, and $f_{2}^{\prime}$, respectively.}\label{tab:decay:ssss}
\tabcolsep=0.15 em
\resizebox{\linewidth}{!}{ 
\begin{tabular}{l|rrrr|rrrrrrr|rrr|rrrrrrrr}
\hline
\hline
State & $\eta\eta$ & $\eta\eta^{\prime}$ & $\eta^{\prime}\eta^{\prime}$ & $\phi\phi$ & $\eta f_{0}$ & $\eta^{\prime}f_{0}$ & $\eta f_{1}$ & $\eta^{\prime}f_{1}$ & $\eta f_{2}^{\prime}$ & $\eta^{\prime}f_{2}^{\prime}$ & $\phi h_{1}$ & $\eta\eta^{\prime}$$\left(2S\right)$ & $\eta^{\prime}\eta^{\prime}$$\left(2S\right)$ & $\phi\phi$$\left(2S\right)$  & $f_{0}f_{0}$ & $f_{0}f_{1}$ & $f_{0}f_{2}^{\prime}$ & $f_{1}f_{1}$ & $f_{1}f_{2}^{\prime}$ & $h_{1}h_{1}$ & $f_{2}^{\prime}f_{2}^{\prime}$ & \tabularnewline
$M_{f_{1}}+M_{f_{2}}$ & 1096 & 1506 & 1916 & 2038 & 1921 & 2331 & 2040 & 2450 & 2065 & 2475 & 2428 & 2167 & 2577 & 2699 & 2746 & 2865 & 2890 & 2984 & 3009 & 2818 & 3034 & \tabularnewline
\hline
$T_{(4s)0^{++}\left(2218\right)\left(1S\right)}$ & 1.6 & 5.5 & 4.4 & 6.9 & - & - & - & - & - & - & - & 0.1 & - & - & - & - & - & - & - & - & - & \tabularnewline
$T_{(4s)0^{++}\left(2440\right)\left(1S\right)}$ & 0.1 & 0.3 & 0.2 & 17.8 & - & - & - & - & - & - & - & o & - & - & - & - & - & - & - & - & - & \tabularnewline
$T_{(4s)2^{++}\left(2378\right)\left(1S\right)}$ & 0.2 & 0.4 & 0.2 & - & - & - & - & - & - & - & - & o & - & - & - & - & - & - & - & - & - & \tabularnewline
\hline
$T_{(4s)0^{++}\left(2798\right)\left(2S\right)}$ & 0.6 & 2.4 & 2.3 & 0.7 & - & - & 0.2 & 0.4 & - & - & 8.8 & 7.0 & 23.9 & 13.1 & 0.1 & - & - & - & - & - & - & \tabularnewline
$T_{(4s)0^{++}\left(2876\right)\left(2S\right)}$ & o & o & o & 21.1 & - & - & 0.3 & 0.5 & - & - & 2.1 & 0.1 & 0.5 & 22.0 & 3.8 & - & - & - & - & 191.9 & - & \tabularnewline
$T_{(4s)0^{++}\left(2954\right)\left(2S\right)}$ & 0.4 & 1.3 & 1.1 & 1.3 & - & - & 0.5 & 0.9 & - & - & 3.2 & 1.3 & 3.4 & 36.8 & 2.1 & - & o & - & - & 241.1 & - & \tabularnewline
$T_{(4s)0^{++}\left(3155\right)\left(2S\right)}$ & o & 0.1 & 0.1 & 1.5 & - & - & 0.1 & 0.3 & - & - & 0.8 & 1.6 & 4.4 & 19.3 & 0.3 & - & 1.1 & 7.3 & 4.6 & 14.8 & 23.0 & \tabularnewline
$T_{(4s)1^{++}\left(2943\right)\left(2S\right)}$ & - & - & - & o & o & o & 0.1 & 0.1 & 0.1 & 0.1 & 0.4 & - & - & 0.2 & - & 0.1 & - & - & - & 0.1 & - & \tabularnewline
$T_{(4s)2^{++}\left(2854\right)\left(2S\right)}$ & 0.2 & 0.4 & 0.2 & 3.7 & - & - & 0.2 & 0.2 & 0.1 & o & 0.3 & 0.1 & 0.1 & 43.2 & o & - & - & - & - & o & - & \tabularnewline
$T_{(4s)2^{++}\left(2981\right)\left(2S\right)}$ & o & 0.1 & o & 6.3 & - & - & 0.2 & 0.5 & 1.5 & 2.6 & 0.6 & 0.1 & 0.1 & 69.6 & 2.8 & 41.8 & 169.8 & - & - & 0.1 & - & \tabularnewline
\hline
$T_{(4s)0^{-+}\left(2481\right)\left(1P\right)}$ & - & - & - & 3.7 & 5.3 & 5.2 & - & - & o & - & 0.6 & - & - & - & - & - & - & - & - & - & - & \tabularnewline
$T_{(4s)0^{-+}\left(2635\right)\left(1P\right)}$ & - & - & - & 14.5 & 2.8 & 6.3 & - & - & 0.6 & 0.2 & 3.4 & - & - & - & - & - & - & - & - & - & - & \tabularnewline
$T_{(4s)0^{-+}\left(2761\right)\left(1P\right)}$ & - & - & - & 2.5 & 9.7 & 22.2 & - & - & 2.7 & 1.9 & 0.4 & - & - & o & - & - & - & - & - & - & - & \tabularnewline
$T_{(4s)1^{-+}\left(2564\right)\left(1P\right)}$ & - & - & - & 0.1 & - & - & 1.3 & 1.9 & 0.2 & o & 1.0 & o & - & - & - & - & - & - & - & - & - & \tabularnewline
$T_{(4s)1^{-+}\left(2632\right)\left(1P\right)}$ & - & - & - & 2.6 & - & - & 0.1 & o & 0.1 & o & 16.6 & - & - & - & - & - & - & - & - & - & - & \tabularnewline
$T_{(4s)1^{-+}\left(2778\right)\left(1P\right)}$ & - & - & - & o & - & - & 4.9 & 10.0 & o & o & o & o & - & - & - & - & - & - & - & - & - & \tabularnewline
$T_{(4s)2^{-+}\left(2537\right)\left(1P\right)}$ & - & - & - & 0.1 & 0.1 & 0.1 & 0.1 & o & 1.3 & 1.9 & 11.1 & - & - & - & - & - & - & - & - & - & - & \tabularnewline
$T_{(4s)2^{-+}\left(2669\right)\left(1P\right)}$ & - & - & - & 0.1 & 0.3 & 0.2 & 0.3 & 0.1 & 2.1 & 3.8 & 14.8 & - & - & - & - & - & - & - & - & - & - & \tabularnewline
$T_{(4s)2^{-+}\left(2837\right)\left(1P\right)}$ & - & - & - & o & 0.7 & 0.6 & 1.0 & 0.8 & 1.0 & 0.5 & 1.6 & - & - & o & - & - & - & - & - & - & - & \tabularnewline
\hline
\hline
State & $\eta\phi$ & $\eta^{\prime}\phi$ & & & $\eta h_{1}$ & $\eta^{\prime}h_{1}$ & $f_{0}\phi$ & $f_{1}\phi$ & $f_{2}^{\prime}\phi$ & & & $\eta\phi$$\left(2S\right)$ & $\eta^{\prime}\phi$$\left(2S\right)$ & $\phi\eta$$\left(2S\right)$ & $f_{0}h_{1}$ & $f_{1}h_{1}$ & $f_{2}^{\prime}h_{1}$ & & & & & \tabularnewline
$M_{f_{1}}+M_{f_{2}}$ & 1567 & 1977 & & & 1957 & 2367 & 2392 & 2511 & 2536 & & & 2228 & 2638 & 2638 & 2782 & 2901 & 2926 & & & & & \tabularnewline
\hline
$T_{(4s)1^{+-}\left(2323\right)\left(1S\right)}$ & 3.3 & 5.0 & & & - & - & - & - & - & & & 0.1 & - & - & - & - & - & & & & & \tabularnewline
\hline
$T_{(4s)0^{+-}\left(2891\right)\left(2S\right)}$ & - & - & & & - & 0.2 & 1.2 & 2.4 & 5.5 & & & - & - & - & - & - & - & & & & & \tabularnewline
$T_{(4s)0^{+-}\left(2967\right)\left(2S\right)}$ & - & - & & & 1.8 & 3.0 & 0.1 & 0.4 & 0.4 & & & - & - & - & - & 0.1 & o & & & & & \tabularnewline
$T_{(4s)1^{+-}\left(2835\right)\left(2S\right)}$ & 3.3 & 6.1 & & & o & 0.1 & 0.1 & 0.5 & o & & & 0.5 & 2.2 & 14.4 & 11.3 & - & - & & & & & \tabularnewline
$T_{(4s)1^{+-}\left(2950\right)\left(2S\right)}$ & 0.2 & 0.5 & & & 0.5 & 0.9 & 0.4 & 1.8 & 5.5 & & & 3.6 & 12.8 & 47.9 & 31.1 & 387.7 & 719.8 & & & & & \tabularnewline
$T_{(4s)2^{+-}\left(2965\right)\left(2S\right)}$ & - & o & & & o & 0.1 & o & 0.1 & 0.3 & & & - & o & o & o & 0.2 & 0.4 & & & & & \tabularnewline
\hline
$T_{(4s)0^{--}\left(2507\right)\left(1P\right)}$ & 2.3 & 3.0 & & & - & - & - & - & - & & & 0.1 & - & - & - & - & - & & & & & \tabularnewline
$T_{(4s)0^{--}\left(2821\right)\left(1P\right)}$ & 0.7 & 1.7 & & & - & - & - & 47.2 & 0.6 & & & 0.4 & 0.3 & o & 0.2 & - & - & & & & & \tabularnewline
$T_{(4s)1^{--}\left(2445\right)\left(1P\right)}$ & 0.1 & o & & & o & o & 8.6 & - & - & & & o & - & - & - & - & - & & & & & \tabularnewline
$T_{(4s)1^{--}\left(2567\right)\left(1P\right)}$ & 2.3 & 3.4 & & & 1.0 & 0.8 & 2.3 & 9.9 & 6.5 & & & o & - & - & - & - & - & & & & & \tabularnewline
$T_{(4s)1^{--}\left(2627\right)\left(1P\right)}$ & o & 0.1 & & & 1.3 & 1.6 & 0.7 & 1.6 & 3.0 & & & o & - & - & - & - & - & & & & & \tabularnewline
$T_{(4s)1^{--}\left(2766\right)\left(1P\right)}$ & 0.2 & 0.2 & & & 0.9 & 0.7 & 27.6 & 0.1 & 0.7 & & & 0.1 & 0.1 & o & - & - & - & & & & & \tabularnewline
$T_{(4s)1^{--}\left(2984\right)\left(1P\right)}$ & 0.5 & 0.7 & & & 7.0 & 13.8 & 0.7 & 9.8 & 25.4 & & & o & o & o & 5.1 & 4.3 & 5.8 & & & & & \tabularnewline
$T_{(4s)2^{--}\left(2446\right)\left(1P\right)}$ & o & o & & & o & - & o & - & - & & & 0.1 & - & - & - & - & - & & & & & \tabularnewline
$T_{(4s)2^{--}\left(2657\right)\left(1P\right)}$ & 0.1 & 0.2 & & & 0.2 & 0.1 & o & 20.4 & 15.8 & & & 0.1 & o & - & - & - & - & & & & & \tabularnewline
$T_{(4s)2^{--}\left(2907\right)\left(1P\right)}$ & 0.7 & 1.2 & & & o & o & 0.8 & 14.5 & 1.6 & & & 0.1 & 0.1 & o & 9.6 & 0.4 & - & & & & & \tabularnewline
$T_{(4s)3^{--}\left(2719\right)\left(1P\right)}$ & 0.1 & 0.1 & & & 0.1 & 0.1 & 0.7 & 1.0 & 64.2 & & & o & - & - & - & - & - & & & & & \tabularnewline
\hline
\hline
\end{tabular}
}
\end{table*}

\section{Results and discussions}\label{sec:results}

The decay results are presented in Table~\ref{tab:decay:ssss}. For the ground state, the spin-spin interactions play a crucial role for the fall-apart decays of a multiquark state. However, for some excited states, the linear confinement potential and/or Coulomb potential also have significant contributions. For most of the $T_{ss\bar{s}\bar{s}}$ states, the width contributed by the fall-apart decays is no more than 50 MeV, while several excited states possess much larger widths due to the opening of specific channels. Based on the mass spectrum and fall-apart decay properties, some discussions about the $T_{ss\bar{s}\bar{s}}$ states are given as follows.

\subsection{$1S$ states}

According to our quark model predictions, there are four $1S$ states $T_{(4s)0^{++}}(2218)$, $T_{(4s)0^{++}}(2440)$,
$T_{(4s)1^{+-}}(2323)$, and $T_{(4s)2^{++}}(2378)$~\cite{Liu:2020lpw}. Their masses are significantly greater than
the $\phi\phi$ threshold, which allows them to decay into the $\eta\eta$, $\eta\eta^{\prime}$, $\eta^{\prime}\eta^{\prime}$,
and/or $\phi\phi$ channels through quark rearrangement.

\subsubsection{$J^{PC}=0^{++}$ state}

The $T_{(4s)0^{++}}(2218)$, as the lowest $T_{ss\bar{s}\bar{s}}$ state, has a fall-apart width of $\Gamma\simeq 19$ MeV, which is mainly contributed by the $\eta\eta^{\prime}$, $\eta^{\prime}\eta^{\prime}$, and $\phi\phi$ channels, and also has sizeable decay rates into the $\eta\eta$ channel. The partial width ratios between these channels are estimated to be
\begin{eqnarray}
\Gamma[\eta\eta]:\Gamma[\eta\eta^{\prime}]:\Gamma[\eta^{\prime}\eta^{\prime}]:\Gamma[\phi\phi]&\simeq &1.0:3.4:2.7:4.3\ .
\end{eqnarray}
While the high-lying $1S$ state $T_{(4s)0^{++}}(2440)$ also has a fall-apart width of $\Gamma\simeq 18$ MeV, which is nearly saturated by the $\phi\phi$ channel, the decay rates into the $\eta\eta^{\prime}$ and $\eta^{\prime}\eta^{\prime}$ channels are estimated to be $\mathcal{O}(1\%)$.

Some evidences of the ground $0^{++}$ $T_{ss\bar{s}\bar{s}}$ states may have been observed in experiments. In Ref.~\cite{Kozhevnikov:2019lmy}, Kozhevnikov carried out a dynamic analysis of the resonance contributions to $J/\psi\to\gamma X\to\gamma\phi\phi$ by using the data from BESIII~\cite{BESIII:2016qzq}. This analysis indicates that there may exist two $0^{++}$ resonances with masses $\sim2.2$~GeV and $\sim2.4$~GeV in the $\phi\phi$ invariant mass spectrum. The two scalar resonances extracted from the data of $J/\psi\to\gamma X\to\gamma\phi\phi$ are likely candidates of the states $T_{(4s)0^{++}}(2218)$ and $T_{(4s)0^{++}}(2440)$ predicted in the quark model. Furthermore, the analysis of the decay data of $B^0\to J/\psi \phi\phi$~\cite{LHCb:2016mkf} also shows an evidence of a scalar resonance around 2.2 GeV in the $\phi\phi$ mass spectrum~\cite{Kozhevnikov:2017nlr}.

It should be mentioned that the evidences of the scalar resonance around 2.2 GeV observed in the $\phi\phi$ channel may be consistent with the resonance $f_{0}(2200)$ listed in RPP~\cite{ParticleDataGroup:2024cfk}. The mass of $f_{0}(2200)$ cannot be explained with a conventional $s\bar{s}$ state, however, is consistent with the ground $T_{ss\bar{s}\bar{s}}$ state predicted in the quark model~\cite{Ebert:2008id}. For the $f_{0}(2200)$ resonance, the $K\bar{K}$ and $\eta\eta$ channels have been observed in experiments. To confirm the nature of $f_{0}(2200)$ and establish the $0^{++}$ $1S$-wave $T_{ss\bar{s}\bar{s}}$ states, more observations of the $\eta\eta^{\prime}$, $\eta^{\prime}\eta^{\prime}$ and $\phi\phi$ channels are expected to be carried out in future experiments.

Finally, it should be mentioned that the full-strange tetraquark states $T_{(4s)0^{++}}(2218)$ and $T_{(4s)0^{++}}(2440)$ are flavor partners of the fully-charmed tetraquark states $T_{(4c)0^{++}}(6455)$ and $T_{(4c)0^{++}}(6550)$ predicted in our previous works~\cite{Liu:2019zuc,liu:2020eha}, respectively. Some evidence of these two scalar fully-charmed tetraquark states may have been observed in recent LHC experiments as well. The $T_{(4c)0^{++}}(6455)$ and $T_{(4c)0^{++}}(6550)$ may be the contributors to the bump structures around $6.4-6.6$ GeV observed in the di-$J/\psi$
invariant mass spectrum at LHC~\cite{CMS:2023owd,LHCb:2020bwg}.

\subsubsection{$J^{PC}=2^{++}$ state}

For the $1S$ state $T_{(4s)2^{++}}(2378)$, the fall-apart width is a very small value of $\Gamma\simeq 1$ MeV. It is interesting to note that the decay channel into $\phi\phi$ is forbidden. The prohibition of the decay of the $1S$ state $T_{(4s)2^{++}}(2378)$ into $\phi\phi$ originates from the exact cancellation among the dynamical transition matrix elements. Specifically, under the configuration of the wavefunction $\psi_{000}^{1S}\chi_{2m_s}^{11}\left|\bar{3}3\right\rangle^{c}$, the color matrix elements exhibit opposite signs, satisfying the identity:
\begin{eqnarray}
\langle\bar{3}_{12}3_{34}|\boldsymbol{\lambda}_{1}\cdot\boldsymbol{\lambda}_{2}|1_{13}1_{24}\rangle & = & \langle\bar{3}_{12}3_{34}|\boldsymbol{\lambda}_{3}\cdot\boldsymbol{\lambda}_{4}|1_{13}1_{24}\rangle\nonumber \\
=-\langle\bar{3}_{12}3_{34}|\boldsymbol{\lambda}_{1}\cdot\boldsymbol{\lambda}_{4}|1_{13}1_{24}\rangle & = & -\langle\bar{3}_{12}3_{34}|\boldsymbol{\lambda}_{2}\cdot\boldsymbol{\lambda}_{3}|1_{13}1_{24}\rangle \ .
\end{eqnarray}
Meanwhile, the associated $1S$ spatial wavefunction ensures that the interaction integrals for all quark pairs are identical. Consequently, the two primary components of the decay amplitude, $\langle B_{13}C_{24} | V_{12}+V_{34} | A \rangle$ and $\langle B_{13}C_{24} | V_{14}+V_{23} | A \rangle$ (or the corresponding terms for the $B_{14}C_{23}$ channel) will cancel each other exactly. This leads to a complete nullification of all potential terms driving the decay—including the confinement, Coulomb, spin-spin, and tensor interactions—when taking the summation of $V_{12}, V_{34}, V_{14},$ and $V_{23}$. This dynamical prohibition, rooted in the symmetry of the wavefunction, also explains why the rearrangement decay widths of $ss\bar{s}\bar{s}$ systems are generally suppressed. The partial widths for the main fall-apart channels $\eta\eta$, $\eta\eta^{\prime}$, and $\eta^{\prime}\eta^{\prime}$ are predicted to be $\mathcal{O}(100)$ keV.

In Refs.~\cite{Ebert:2008id,Lu:2019ira}, the tensor resonances $f_{2}(2300)$ and $f_{2}(2340)$ listed in RPP~\cite{ParticleDataGroup:2024cfk} were suggested to be candidates of the $1S$ state $T_{ss\bar{s}\bar{s}}$ with $J^{PC}=2^{++}$ according to the mass analysis. Both the $f_{2}(2300)$ and $f_{2}(2340)$ have been seen in the $\phi\phi$ channel with a broad width of several hundred MeV. While the $f_{2}(2340)$ has been seen in the $\eta\eta^{\prime}$, $\eta^{\prime}\eta^{\prime}$ and $\phi\phi$ channels with a broad width of $\Gamma\simeq 300$ MeV.

However, if assigning $f_{2}(2300)$ or $f_{2}(2340)$ as the $1S$ state $T_{(4s)2^{++}}(2378)$, we will face two challenges: (i) the observed $\phi\phi$ channel cannot be explained. According to our analysis, the $\phi\phi$ channel is forbidden; (ii) the broad width cannot be well understood. The predicted partial widths for the $\eta\eta^{\prime}$ and $\eta^{\prime}\eta^{\prime}$ channels are $\mathcal{O}(100)$ keV, which is negligibly small compared with the measured widths of $f_{2}(2300)$ and $f_{2}(2340)$.
It should be mentioned that from the point of view of mass, the $f_{2}(2300)$/$f_{2}(2340)$ may
be consistent with a tensor glueball predicted within Lattice QCD~\cite{Yang:2013xba,Chen:2005mg}.

\subsubsection{ $J^{PC}=1^{+-}$ state and $X(2300)$ }

The $1S$ state $T_{(4s)1^{+-}}(2323)$ has a narrow fall-apart decay width of $\Gamma\simeq 8$ MeV, which is saturated by the $\eta\phi$ and $\eta^{\prime}\phi$ channels. The partial width ratio between these two channels is predicted to be 
\begin{eqnarray}
 \Gamma[\eta\phi]:\Gamma[\eta^{\prime}\phi]\simeq 1.0:1.5\ .
\end{eqnarray}

Recently, a new $1^{+-}$ state $X(2300)$ with a width of $\Gamma_{exp}=89\pm15\pm26$ MeV was observed in the $\phi\eta$ and $\phi\eta^{\prime}$ invariant mass spectra by using the $\psi(3686)\to \phi\eta\eta^{\prime}$ decay at BESIII~\cite{BESIII:2024nhv}. If assigning $X(2300)$ as the conventional $s\bar{s}$ state $h_1(3P)$, the observed mass is notably smaller than the quark model predictions, while the measured width is too broad compared to the theoretical expectation~\cite{Wang:2024lba}. However, it is interesting to find that the observed mass, decay channels, and quantum numbers of the new resonance $X(2300)$ favor the assignment of the fully-strange tetraquark state $T_{(4s)1^{+-}}(2323)$ predicted within the quark model~\cite{Liu:2020lpw}.

It should be pointed out that if the $X(2300)$ is a $T_{ss\bar{s}\bar{s}}$ state indeed, the two fall-apart channels $\phi\eta$ and $\phi\eta^{\prime}$ cannot saturate its decay width, their branching fractions are estimated to be $\sim 7\%$ and $\sim 11\%$, respectively. The $X(2300)$ may have large potentials to be observed in the other channels, such as $\bar{K}K^*+c.c. $.

Finally, it should be mentioned that if $X(2300)$ corresponds to the low-lying fully-strange tetraquark state with $J^{PC}=1^{+-}$ indeed, its partner in the fully-charmed sector $T_{(4c)1^{+-}}(6500)$ predicted in the quark model~\cite{Liu:2019zuc} may have a large potential to be observed in the $\eta_c J/\psi$ channel.

\subsection{$2S$ states}

There are twelve $2S$ states predicted in the quark model. Their
masses scatter in the range of $\sim2.8-3.2$ GeV~\cite{Liu:2020lpw},
which are significantly greater than the $\phi\phi(1680)$ threshold.
Thus, many fall-apart decay channels as listed in Table~\ref{tab:decay:ssss} are opened for these states. 

\subsubsection{$J^{PC}=0^{++}$ states}

According to predictions, there are four $2S$ states
with $J^{PC}=0^{++}$: $T_{(4s)0^{++}}(2798)$, $T_{(4s)0^{++}}(2876)$,
$T_{(4s)0^{++}}(2954)$, and $T_{(4s)0^{++}}(3155)$~\cite{Liu:2020lpw}.

The low-lying $2S$ state $T_{(4s)0^{++}}(2798)$ has a fall-apart
decay width of $\Gamma\simeq 60$ MeV, and mainly decays into the $\eta^{\prime}\eta^{\prime}(2S)$
and $\phi\phi(1680)$ channels with branching fractions $\sim40\%$ and $\sim22\%$, respectively.

While for the high-lying $2S$ state $T_{(4s)0^{++}}(3155)$, the fall-apart
decay width is estimated to be $\Gamma \simeq 78$ MeV. This state has a large decay rate
into the $\phi\phi(1680)$ channel with a branching fraction of about $25\%$.

Both the $T_{(4s)0^{++}}(2876)$ and $T_{(4s)0^{++}}(2954)$ states may be too broad to be observed in experiments.
Their partial widths into the $h_{1}(1415)h_{1}(1415)$ channel are estimated to be $\sim 200-250$ MeV due
to a large overlap of the spatial wave functions of initial and final states.
Except for the dominant $h_{1}(1415)h_{1}(1415)$ channel, they also have significant decay rates
into the $\phi\phi(1680)$ channel with a partial width of $\mathcal{O}(10)$ MeV.

To search for the $0^{++}$ $2S$ $T_{ss\bar{s}\bar{s}}$ states,
the $\phi\phi(1680)$ channel is worth to observing in future experiments.

\subsubsection{$J^{PC}=2^{++}$ states}

According to our quark model predictions, there are two $2S$ states
with $J^{PC}=2^{++}$: $T_{(4s)2^{++}}(2854)$ and $T_{(4s)2^{++}}(2981)$~\cite{Liu:2020lpw}.

The low-lying state $T_{(4s)2^{++}}(2854)$ has a fall-apart width
of $\Gamma\sim 49$ MeV, which is mainly contributed by the $\phi\phi(1680)$
channel with a branching fraction of $\sim89\%$. While the partial
widths for the $\eta\eta$, $\eta\eta^{\prime}$, $\eta^{\prime}\eta^{\prime}$, $\eta\eta(2S)$
, and $\eta^{\prime}\eta(2S)$ channels are predicted to be $\mathcal{O}(100)$
keV.

The high-lying $2S$ state $T_{(4s)2^{++}}(2981)$ may be a broad state with a very large fall-apart
decay width of $\Gamma\simeq 296$ MeV. It is mainly contributed
by the $f_0(1370)f_2'(1525)$ channel, while the contributions from the $\phi\phi$ and $\phi\phi(1680)$ channels are also sizeable.
The partial width ratios between these main channels are predicted to be
\begin{eqnarray}
\Gamma[\phi\phi]:\Gamma[\phi\phi(1680)]:\Gamma[f_0(1370)f_2'(1525)]\simeq1:11:27\ .
\end{eqnarray}

To establish the $2^{++}$ $2S$-wave $T_{ss\bar{s}\bar{s}}$ states, the $\phi\phi(1680)$
invariant mass spectrum around the $2.8-3.0$ GeV region is worth
observing in future experiments.

\subsubsection{$J^{PC}=1^{+-}$ states}

According to our quark model predictions, there are two $2S$ states
with $J^{PC}=1^{+-}$: $T_{(4s)1^{+-}}(2835)$, and $T_{(4s)1^{+-}}(2950)$~\cite{Liu:2020lpw}.

For the $T_{(4s)1^{+-}}(2835)$ state, the fall-apart width is estimated
to be $\Gamma\simeq 38$ MeV. This state has large decay rates into
the $\eta^{\prime}\phi$ and $\phi\eta^{\prime}(2S)$ channels. The partial
width ratio between these two main channels is predicted to be
\begin{eqnarray}
\Gamma[\eta^{\prime}\phi]:\Gamma[\phi\eta^{\prime}(2S)]\simeq1.0:2.4\ .
\end{eqnarray}
The $T_{(4s)1^{+-}}(2835)$ also has sizeable decay rates into the
$\eta\phi$ and $\eta^{\prime}\phi(1680)$ channels with a partial
width in the range of $\sim4-7$ MeV. To search for the $T_{(4s)1^{+-}}(2835)$ state,
the $\eta^{\prime}\phi$ and $\eta^{\prime}\phi(1680)$ channels are worth
observing in future experiments.

The high-lying state $T_{(4s)1^{+-}}(2950)$ may be too broad to be observed in experiments.
The partial widths of $f_{1}h_{1}(1415)$ and $f_{2}^{\prime}(1525)h_{1}(1415)$ are estimated
to be very large values of $\sim 388$ MeV and $\sim 720$ MeV, respectively.

\subsubsection{$J^{PC}=0^{+-},2^{+-}$ states}

According to quark model predictions, there are two $0^{+-}$
states $T_{(4s)0^{+-}}(2891)$ and $T_{(4s)0^{+-}}(2967)$, and one
$2^{+-}$ state $T_{(4s)2^{+-}}(2965)$~\cite{Liu:2020lpw}.

For the $T_{(4s)0^{+-}}(2891)$ state, the fall-apart width is estimated to be
$\Gamma\simeq 10$ MeV. This state may have large
decay rates into the $f_{0}(1370)\phi$, $f_{1}\phi$, and $f_{2}^{\prime}(1525)\phi$
channels with partial width ratios
\begin{eqnarray}
\Gamma[f_{0}(1370)\phi]:\Gamma[f_{1}\phi]:\Gamma[f_{2}^{\prime}(1525)\phi]
\simeq1.0:2.1:4.8\ .
\end{eqnarray}

The high-lying $0^{+-}$ state $T_{(4s)0^{+-}}(2967)$ also has a
narrow the fall-apart width of $\Gamma\simeq 11$ MeV, which is mainly
contributed by the $\eta h_{1}(1415)$ and $\eta^{\prime}h_{1}(1415)$
channels with branching fractions $\sim31\%$ and $\sim54\%$, respectively.

The $2^{+-}$ state $T_{(4s)2^{+-}}(2965)$ has a very narrow fall-apart
width of $\mathcal{O}(100)$ keV, which is mainly contributed by the
$f_{2}^{\prime}(1525)\phi$, $f_{2}^{\prime}(1525)h_1(1415)$, and
$f_{1}h_1(1415)$ with comparable decay rates.

To search for the $2S$-wave states with $J^{PC}=0^{+-}$ and $2^{+-}$,
the $\eta^{(\prime)}h_{1}(1415)$ and $\phi f_{2}^{\prime}(1525)$
channels are worth observing in future experiments.

\subsection{$1P$ states}

There are twenty $1P$ states predicted in the quark model. Their masses scatter in the range of $\sim2.4-3.0$ GeV~\cite{Liu:2020lpw}.
The fall-apart partial decay widths are evaluated and presented in Table~\ref{tab:decay:ssss}.

\subsubsection{$J^{PC}=0^{-+}$ states and $X(2500)$}

According to quark model predictions, there are three $1P$-wave states with $0^{-+}$: $T_{(4s)0^{-+}}(2481)$, $T_{(4s)0^{-+}}(2635)$, and $T_{(4s)0^{-+}}(2761)$~\cite{Liu:2020lpw}. 

The low-lying $0^{-+}$ state $T_{(4s)0^{-+}}(2481)$ has a relatively narrow fall-apart width of $\Gamma\simeq 15$ MeV. The main fall-apart decay channels are $\phi \phi$, $\eta f_{0}(1370)$, and $\eta^{\prime}f_{0}(1370)$. The partial width ratios are predicted to be
\begin{eqnarray}
 \Gamma[\phi\phi]:\eta f_{0}(1370):\Gamma[\eta^{\prime}f_{0}(1370)]\simeq 1.0:1.4:1.4\ .
\end{eqnarray}

For the middle mass state $T_{(4s)0^{-+}}(2635)$, the fall-apart width is estimated to be $\Gamma\simeq 28$ MeV, which is mainly contributed by the $\phi \phi$ channel with a branching fraction of $\sim 52\%$.

While for the high-lying state $T_{(4s)0^{-+}}(2761)$, the fall-apart width is estimated to be $\Gamma\simeq 39$ MeV, which mainly contributed by the $\eta^{\prime}f_{0}(1370)$ and $\eta f_{0}(1370)$ channels with branching fractions $\sim25\%$, $\sim57\%$, respectively. There is a sizeable decay rate into the $\phi\phi$ channel with a branching fraction $\sim6\%$.

Some evidence of $T_{(4s)0^{-+}}(2481)$ may have been observed in experiments. In 2016, the BESIII collaboration observed a new resonance $X(2500)$ with a broad width of $\Gamma\simeq 230$ MeV in the $\phi\phi$ invariant mass spectrum from the $J/\psi\to\gamma\phi\phi$ process~\cite{BESIII:2016qzq}. The spin-parity numbers seem to more favor $0^{-+}$. The observed mass, spin-parity numbers, 
and decay mode of $X(2500)$ are consistent with those of $T_{(4s)0^{-+}}(2481)$ predicted in theory. The $X(2500)$ resonance is also suggested as a $0^{-+}$ $T_{ss\bar{s}\bar{s}}$ candidate based on the mass analyses within the QCD sum rules~\cite{Dong:2020okt,Su:2022eun} and the relativized quark model~\cite{Lu:2019ira}. However, the $X(2500)$ was explained as a conventional $s\bar{s}$ state in the literature~\cite{Pan:2016bac,Xue:2018jvi,Wang:2020due,Li:2020xzs}. To confirm the nature of $X(2500)$, more studies are needed in both theory and experiments.

Finally, it should be mentioned that the $X(2370)$ listed in RPP~\cite{ParticleDataGroup:2022pth} is explained as a $J^{PC}=0^{-+}$ $T_{ss\bar{s}\bar{s}}$ state in the literature~\cite{Dong:2020okt,Su:2022eun}. This resonance was observed in the $\pi\pi\eta^{\prime}$ and $K\bar{K}\eta^{\prime}$ invariant mass spectra via the $J/\psi$ decays at BESIII~\cite{BESIII:2010gmv,BESIII:2019wkp}. Recently, the spin-parity numbers are determined to be $0^{-+}$ by the BESIII Collaboration~\cite{BESIII:2023wfi}. Although the quantum numbers and decay modes are consistent with the $0^{-+}$ $T_{ss\bar{s}\bar{s}}$ states, the measured mass of $X(2370)$ is too low to be comparable with our quark model predictions. The $X(2370)$ may favor a glueball-like particle~\cite{Huang:2025pyv,She:2024ewy,Cao:2024mfn} as that predicted by the Lattice QCD~\cite{Gui:2019dtm,Chen:2005mg}.

\subsubsection{$J^{PC}=1^{--}$ states}

According to our quark model predictions, there are five $1P$-wave states with $1^{--}$: $T_{(4s)1^{--}}(2455)$, $T_{(4s)1^{--}}(2567)$, $T_{(4s)1^{--}}(2627)$, $T_{(4s)1^{--}}(2766)$, and $T_{(4s)1^{--}}(2984)$~\cite{Liu:2020lpw}. The fall-apart partial decay widths are presented in Table~\ref{tab:decay:ssss}. 

The low-lying $1^{--}$ state $T_{(4s)1^{--}}(2455)$ has a fall-apart decay width of $\Gamma\simeq 9$ MeV, which is nearly saturated by the $\phi f_{0}(1370)$ channel.

The second $1^{--}$ state $T_{(4s)1^{--}}(2567)$ has a fall-apart decay width of $\Gamma\simeq 26$ MeV. This state has large decay rates into the $\eta\phi$, $\eta^{\prime}\phi$, $\phi f_{0}(1370)$, $\phi f_{1}(1420)$, and $\phi f_{2}^{\prime}(1525)$ channels. The partial width ratios between these channels are predicted to be
\begin{eqnarray}
&\Gamma[\eta\phi]:\Gamma[\eta^{\prime}\phi]:\Gamma[\phi f_{0}(1370)]:\Gamma[\phi f_{1}(1420)]:\Gamma[\phi f_{2}^{\prime}(1525)]\nonumber \\
&\simeq1.0:1.5:1.0:4.3:2.8\ .
\end{eqnarray}

The third $1^{--}$ state $T_{(4s)1^{--}}(2627)$ has a relatively narrow fall-apart width of $\Gamma\simeq 8$ MeV. The partial widths for the main fall-apart channels, $\eta^{\prime} h_{1}(1415)$, $\phi f_{1}(1420)$, and $\phi f_{2}^{\prime}(1525)$, are estimated in the range of $\sim 1-3$ MeV.

The fourth $1^{--}$ state $T_{(4s)1^{--}}(2766)$ has a fall-apart width of $\Gamma\simeq 30$ MeV, which is mainly contributed by the $\phi f_{0}(1370)$ channel with a branching fraction $\sim91\%$.

The high-lying $1^{--}$ state $T_{(4s)1^{--}}(2984)$ has a broad fall-apart decay width of $\Gamma\simeq 72$ MeV, which is mainly contributed by the $\eta^{(\prime)}h_1(1415)$, $\phi f_{1}(1420)$, and $\phi f_{2}^{\prime}(1525)$ channels. The partial width ratios between these channels are predicted to be
\begin{eqnarray}
 \Gamma[\eta h_1(1415)]:\Gamma[\eta^{\prime}h_1(1415)]:\Gamma[\phi f_{1}(1420)]\nonumber\\:\Gamma[\phi f_{2}^{\prime}(1525)]
 \simeq 1.0 : 2.0 : 1.4 : 3.6\ .
\end{eqnarray}

To search for the $1P$-wave $T_{ss\bar{s}\bar{s}}$ states with $J^{PC}=1^{--}$, the $\phi f_{0}(1370)$, $\phi f_{1}(1420)$, and $\phi f_{2}^{\prime}(1525)$ channels are suggested to be observed in future experiments.

Some evidence of the $1P$-wave $T_{ss\bar{s}\bar{s}}$ states with $1^{--}$ may have been observed in experiments. For example, in the $\phi f_{0}(980)$ invariant mass spectrum from the Babar~\cite{BaBar:2007ptr}, Belle~\cite{Belle:2008kuo,Shen:2009mr}, and BESIII~\cite{BESIII:2021lho} observations, there exists a vague peak structure around 2.4 GeV (denoted by $X(2400)$). This structure may be contributed by the low-lying $1^{--}$ state $T_{(4s)1^{--}}(2455)$. The $X(2400)$ is also explained as the $J^{PC}=1^{--}$ $T_{ss\bar{s}\bar{s}}$ state based on the analysis of QCD sum rules~\cite{Jiang:2023atq,Su:2022eun}. To confirm the nature of $X(2400)$, some other channels, such as $\phi f_{0}(1370)$ and $\phi f_{2}^{\prime}(1525)$, are worth observing in future experiments.

Finally, it should be mentioned that the well-known vector resonance $\phi(2170)$ listed in RPP~\cite{ParticleDataGroup:2022pth} is explained as a $1^{--}$ $T_{ss\bar{s}\bar{s}}$ state based on the mass analysis of QCD sum rules~\cite{Wang:2006ri,Chen:2008ej,Chen:2018kuu,Wang:2019nln,Cui:2019roq,Su:2022eun,Jiang:2023atq} and the flux-tube model~\cite{Deng:2010zzd}. However, according to our quark model predictions, we find that the measured mass of $\phi(2170)$ is about 270~MeV smaller than the lowest $J^{PC}=1^{--}$ state $T_{(4s)1^{--}}(2455)$. The $\phi(2170)$ may favor other assignments, such as a conventional $s\bar{s}$ state~\cite{Ding:2007pc,Wang:2012wa,Pang:2019ttv}, a hidden-strange baryonium or hexaquark state~\cite{Abud:2009rk,Zhao:2013ffn,Deng:2013aca,Dong:2017rmg,Cao:2018kos,Zhu:2019ibc}, a dynamically generated state~\cite{Napsuciale:2007wp,GomezAvila:2007ru,MartinezTorres:2008gy,AlvarezRuso:2009xn,Coito:2009na,Xie:2010ig}, or a strangeonium hybrid state~\cite{Ding:2006ya}, as suggested in the literature.

\subsubsection{$P$-wave states with other quantum numbers}

There is no experimental evidence for the other $1P$-wave $T_{ss\bar{s}\bar{s}}$ states with quantum numbers $1^{-+}$, $2^{-+}$, $0^{--}$, $2^{--}$, and $3^{--}$. The fall-apart decay properties have been evaluated and presented in Table~\ref{tab:decay:ssss}.

These states may have large potentials to be observed in their main fall-apart decay channels. For example, the $T_{(4s)0^{--}}(2507)$ state with $J^{PC}=0^{--}$ and the $T_{(4s)1^{-+}}(2632)$ state with $J^{PC}=1^{-+}$ predicted in the quark model may be likely to be established in the $\eta^{(\prime)}\phi$ and $\phi h_1(1415)$ channels, respectively, while the two $2^{--}$ states $T_{(4s)2^{--}}(2657)$ and $T_{(4s)2^{--}}(2907)$, and the $3^{--}$ state $T_{(4s)3^{--}}(2719)$ may be first observed in the $\phi f_{2}^{\prime}(1525)$ channel.

\section{summary}\label{sec:sum}

Based on the mass spectra of the $1S$-, $1P$-, and $2S$-wave fully-strange tetraquark states obtained in the previous work of our group~\cite{Liu:2020lpw}, we continue to study their fall-apart decays within a quark-exchange model. Most of the fully-strange tetraquark states have a relatively narrow fall-apart decay width of $\mathcal{O}(10)$ MeV. Our theoretical predictions indicates that some evidence of fully-strange tetraquark states may have been observed in the previous experiments. 

Some key points are emphasized as follows:
\begin{itemize}
\item The axial-vector state $X(2300)$ newly observed at BESIII favors the low-lying $1S$-wave $T_{ss\bar{s}\bar{s}}$ state with $J^{PC}=1^{+-}$, $T_{(4s)1^{+-}}(2323)$. 
\item The $X(2500)$ observed in the $\phi\phi$ invariant mass spectrum of the $J/\psi\to\gamma\phi\phi$ process at BESIII may be a candidate of the low-lying $1P$-wave $T_{ss\bar{s}\bar{s}}$ state with $J^{PC}=0^{-+}$, $T_{(4s)0^{-+}}(2481)$. 
\item Evidence of the $1S$-wave $T_{ss\bar{s}\bar{s}}$ states with $J^{PC}=0^{++}$ may have been observed in the $\phi\phi$ invariant mass spectrum around $2.2$ GeV and $2.4$ GeV.
\item Evidence of a $1P$-wave $T_{ss\bar{s}\bar{s}}$ state with $J^{PC}=1^{--}$ may have been observed in the $\phi f_0(980)$ invariant mass spectrum around $2.4$ GeV.
\item The $1S$ state $T_{(4s)2^{++}}(2378)$ is prohibited from decaying into $\phi\phi$ due to an exact cancellation mechanism. The $\phi(2170)$, $X(2370)$, $f_{2}(2300)$, and $f_{2}(2340)$ resonances listed in RPP cannot be explained with $T_{ss\bar{s}\bar{s}}$ states.
\end{itemize}

We show that more $S$-wave $T_{ss\bar{s}\bar{s}}$ states can be searched for in the $\phi\phi$ and $\phi\phi(1680)$ channels. It is also strongly recommended that more $1P$-wave states can be investigated in the $\eta^{(\prime)}\phi$, $\eta^{(\prime)}h_1(1415)$ and $\phi f_2^{\prime}(1525)$ channels in future experiments at BESPCII/BESIII and Belle-II. It is noteworthy that some of these states with strong $S$-wave couplings to the nearby open threshold, may serve as a short-distance component in the formation of the physical states of which the long-distance components are hadronic molecules. In the future studies a combined analysis would allow us to have a better understanding of the unerlying dynamics. 

\begin{acknowledgments}

This work is supported by the National Natural Science Foundation of China (Grants No.12235018, No.12175065, No.E411645Z10, No.1A2024000016). 

\end{acknowledgments}

\bibliography{refs}

\end{document}